\documentclass[english]{bullsrsl}

\usepackage[latin1]{inputenc}

\usepackage[T1]{fontenc}

\usepackage{natbib}          

\usepackage{graphicx}
\usepackage{rotating}

\begin{document}
\title{The Polstar UV Spectropolarimetry Mission}




\author[affil=1, corresponding]{Richard}{Ignace}
\author[affil=2]{Paul}{Scowen}


\affiliation[1]{East Tennessee State University, Johnson City, TN 37615, USA}
\affiliation[2]{Code 667, Exoplanets \& Stellar Astrophysics Lab,
	 NASA/GSFC, USA}


\correspondance{ignace@etsu.edu}


\maketitle

\begin{abstract}

The {\em Polstar} small explorer concept is for an ultraviolet
(UV) spectropolarimetry space telescope mission with a focus on massive
star astrophysics.  The instrument waveband will be from 115~nm --
286~nm for spectroscopy and 122~nm -- 286~nm for polarimetry.  All
4 Stokes parameters, IQUV, will be measured at a resolving power
of $R=20,000$ (15 km/s velocity resolution).  The telescope aperture
will be 40~cm with an effective area of about 22~cm$^2$ at a reference
wavelength of 150~nm.  The thrust of the science goals will be to
determine the astrophysics of angular momentum exchange and transport, and
consequences for massive star properties and
evolution.  This includes the effects of rapid
to critical rotation for individual stars (magnetic and non-magnetic),
and the effects of mass transfer for massive binaries, including
identification of stripped core stars.  If selected by the NASA/SMEX
program, {\em Polstar} would launch around 2031 and observe $\sim 300$
stars to achieve science goals.  The mission will include a
Guest Observer program to advance discovery in other areas of
astrophysics.

\end{abstract}

\section{Introduction}

Massive stars are hot, UV-bright, luminous stars
with short lifetimes that terminate in dramatic
style leaving behind compact object remnants \citep{2012ARA&A..50..107L}.
Moreover, they are central in the cosmic narrative for so many other
areas of astrophysics.  Massive stars are important for galaxy evolution as
sites for nucleosynthesis and spewing metals throughout galaxies,
making Earth-like worlds possible \citep{2013ARA&A..51..457N}.  The
story of massive stars further intersects with gravitational wave
detections involving stellar mass black holes
\citep[e.g.,][]{2010ApJ...714.1217B}.

However, a full story of massive stars themselves continues to be a
topic of debate and investigation
\citep[e.g.,][]{2022ARA&A..60..203V}.  We understand now that massive
stars are born into binaries, triples, or higher multiples
\citep{2012Sci...337..444S, 2023A&A...678A..60K}.  Relative to the
critical speed of breakup at the stellar equator, 
nearly all are rapidly rotating (above $\sim$30\%), and some are
rotating near critical (above 90\%).  Those that are slowly
rotating are likely magnetic \citep{2009MNRAS.392.1022U}.  Binarity
combined with the incidence of runaway stars \citep{2019A&A...624A..66R}
and high rotation rates \citep{1996ApJ...463..737P, 2010ApJ...722..605H}
makes clear that massive stars can undergo substantive interactions,
such as mass transfer, common envelope, and even mergers
\citep[e.g.,][]{2024arXiv240711680N}.  All of these involve the
exchange of angular momentum, including orbital evolution and stellar
spin up \citep{1992ApJ...391..246P, 2020A&A...640A..16T}.  It remains
unclear how much spin-up occurs, how angular momentum is transported
throughout stellar interiors, and how much is conserved or lost to
the system during mass exchange.

Although stellar evolution is driven by gravity and therefore by mass,
issues of rotation are consequential.  The {\em Polstar}
mission will bring ultraviolet (UV) spectropolarimetry to bear on
outstanding questions concerning angular momentum exchange and
internal transport -- UV because this where massive stars have
peak brightnesses, spectroscopy because of the strong
resonance lines and wide range of ion species to be found in the
UV, and polarimetry because it probes geometry
\citep[e.g.,][]{2010stpo.book.....C}.  The latter is the novel
aspect of {\em Polstar}.  The stars are unresolved, and the issues
of fast rotation, binarity, and magnetism are intrinsically
aspherical in nature.  Polarization is sensitive to deviations from
sphericity and will provide new diagnostics for advancing our understanding
of massive stars.  This contribution provides an overview of UV spectral
and polarimetric diagnostics and a summary of the
{\em Polstar} design and capabilities.

\section{Spectropolarimetric Diagnostics for Massive Stars}

Polarization is measured using Stokes Parameters I, Q, U, and V.  
Stokes-I refers to the total measure of light.  Stokes-V is
a measure of circular polarization.  Two parameters are
needed for linear polarization, Q and U.  It is typical to
measure polarization in relative terms.  For Stokes fluxes
$F_I$, $F_Q$, and $F_U$, relative parameters are
$q=F_Q/F_I$, and $u=F_U/F_I$.  Then the degree of polarization
$p$ and polarization position angle $\psi_{\rm p}$ become

\[ p = \sqrt{q^2+u^2},~{\rm and}~\tan 2\psi_{\rm p} = u/q. \]

\noindent For applications to massive stars at UV and visible
wavelengths, polarizations are typically small, with 1\% polarization
being fairly sizable.  It is often (but not always) convenient to
treat linear and circular polarizations separately, with circular
polarization $p_V = F_V/F_I$.

\subsection{Continuum Polarization}

\subsubsection{Polar Brightening and Gravity Darkening}

\begin{figure}[t]
\centering
\includegraphics[width=4.5in,angle=-90]{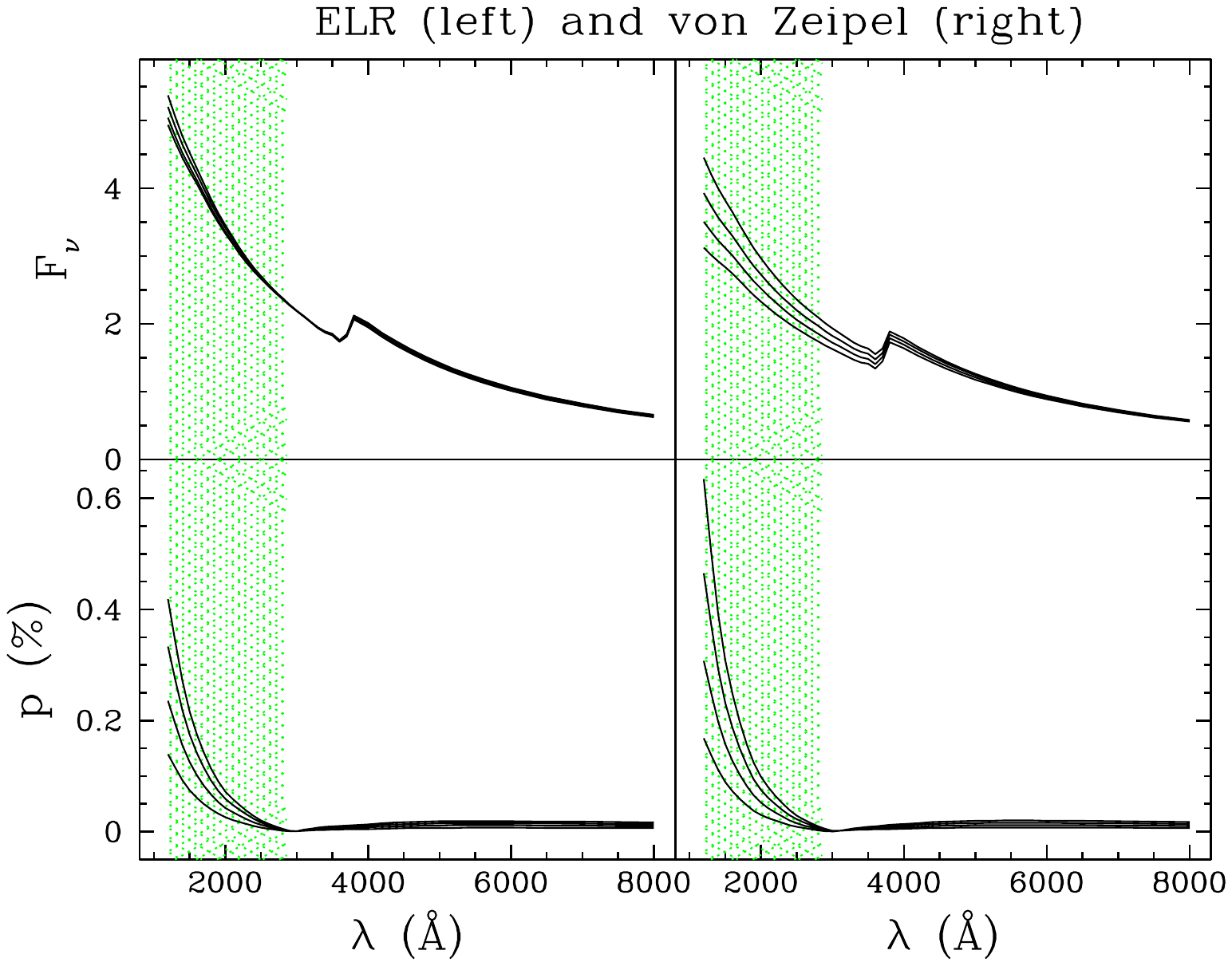}
\begin{minipage}{6in}
\centering
\caption{Net polarization produced from a gravity darkened
B1V~star ($T_{\rm eff}=25,700$~K when not rotating)
at $\omega=0.80$ (lowest polarization), 0.90, 0.95, and 0.975 (highest
polarization) for the star viewed edge-on.
Top is the flux distribution; bottom is polarization.
Left is for ELR; right is for von Zeipel.  Green shading is the
{\em Polstar} waveband.
\label{f1}}
\end{minipage}
\end{figure}

Critical rotation is typically defined as $v_{\rm crit} = \sqrt{GM/R}$, for
stellar mass $M$, stellar radius $R$, and gravitational constant
$G$.  When the equator of a star rotates at this critical speed,
gas is no longer bound.  However, other effects also come into play.
For example, at high rotations the star is distorted from spherical
(``oblateness''), making radius a function of latitude.  Additionally,
the surface temperature becomes a function of latitude, being hotter
toward the pole (``polar brightening'') and lower at the equator
(``gravity darkening''), relative to a non-rotating star
\citep[][(ELR)]{1924MNRAS..84..665V, 2011A&A...533A..43E}.  
As gravity drops at the equator, a star can become closer to the
Eddington limit.  
Consequently, gravity can be reduced at the equator by rotation and
radiation both \citep{2000A&A...361..159M}.  

Let $\omega$ be a dimensionless parameter for rotation relative to
critical (now defined to account for oblateness and radiation as
well), with $\omega<1$ for gas at the equator to remain bound.
The effects of near-critical rotation tend to scale as $\sqrt{1-
\omega^2}$, meaning that even up to 50\% of critical, geometrical
distortions are only at the level of 10\%.  It is rotation
rates above about 80\% where rotational influences can become 
dominant.  One consequence for {\em Polstar} is a net
polarization from the stellar atmosphere \citep{1968ApJ...151.1051H,
1991ApJS...77..541C}.  Towards the FUV, the strongly aspherical
star and temperature distribution can lead to significant net
polarizations of around 1\% and more (see Fig.~\ref{f1} for the
conservative case of a B1V star).  Ground-based
efforts have made sensitive detections of this effect in a limited
number of bright stars at much lower levels in the optical band.
Gravity darkening makes specific predictions for 
polarization with a dramatic rise toward the FUV.
Measurements of that distribution is one of the few
ways to ascertain precisely how close to critical the fastest
rotating stars get.


\subsubsection{Circumstellar Scattering}

For hot massive stars, H is highly ionized (and for Wolf-Rayet
[WR] stars, He
is ionized).  The dominant polarigenic continuum opacity is 
electron scattering.  The optical depth through a circumstellar
medium is of the form

\begin{equation}
\tau_{\rm e} = \int n_{\rm e}\,\sigma_T\,dr,
\end{equation}

\noindent where $\tau_{\rm e}$ is the electron optical depth along a
radial integration of the electron density $n_{\rm e}$, and $\sigma_T$
is the Thomson scattering cross-section.  The latter is fairly low
at $\sigma_T=6.656\times 10^{-25}$~cm$^2$.  The electron optical
depth is thus column depth multiplied by the cross-section.  The column
depth is approximately a product of the base density of the medium
(e.g., the wind base or the inner edge of a disk) multiplied by a
characteristic length scale (generally the stellar radius).  For a spherical
wind, the base density scales roughly as $n_0\sim
\dot{M}/4\pi\,\mu_{\rm e}\,m_H\, R^2\,v_\infty$, for $R$ the radius,
$m_H$ the mass of H, $v_\infty$ the wind terminal speed, and $\mu_{\rm
e}$ the mean molecular weight per free electron.  For O and B stars,
the latter will be of order unity; for WRs, it may be around 2 or
3.  The expression does not account for integrating through the wind
acceleration zone, which may increase the density scale by a factor
of a few.  Rough estimates for electron optical depths of 
{\em Polstar} target categories range from negligible for B~winds to
a few for WR~winds.  However, the disks of Be~stars and of interacting
binaries can have optical depths of order unity or more.


\subsection{Line Polarization Effects}

\subsubsection{\"{O}hman Effect}

The \"{O}hman effect refers to a polarization change across rotationally
broadened photospheric lines \citep{1946ApJ...104..460O,
2024arXiv240711352B}.  The isovelocity zones of a rotating star are
vertical strips, ranging from $-v\sin i$ to $+v\sin i$.  For an
$x-y$ sky coordinate system, with $y$ the projection of the
star's rotation axis, the strips are parallel to $y$.
A velocity bin of $\Delta v$ maps to a strip of width $\Delta x$.
This is approximate since thermal/turbulent broadening causes 
strips not to have geometrically sharp edges.

To illustrate, ignore rotational distortion and assume the star is
spherical.  It will have a centro-symmetric pattern of limb
polarization, which is also wavelength-dependent (generally larger
toward UV wavelengths).  Thus rotational broadening creates a
breaking of symmetry, since a strip at $x$, corresponding to
a line-of-sight velocity shift $v_{\rm z}$, is relatively more absorbing
of polarized light (assuming a pure absorption line).  The
result is that the remainder of the centro-symmetric pattern no
longer cancels.

The Stokes-Q polarization shows a triple peak pattern, with two
negative polarizations for the line wings, and a positive one at
line center.  
By contrast the Stokes-U polarization is antisymmetric
and zero at line center.  The polarization tends to be weaker, and
weaker in $U$ than $Q$.  The detailed profile shapes are sensitive
to the limb polarization profile, from center to limb, so can be
used to probe stellar atmosphere models.

\begin{figure}[t]
\centering
\includegraphics[width=3.25in]{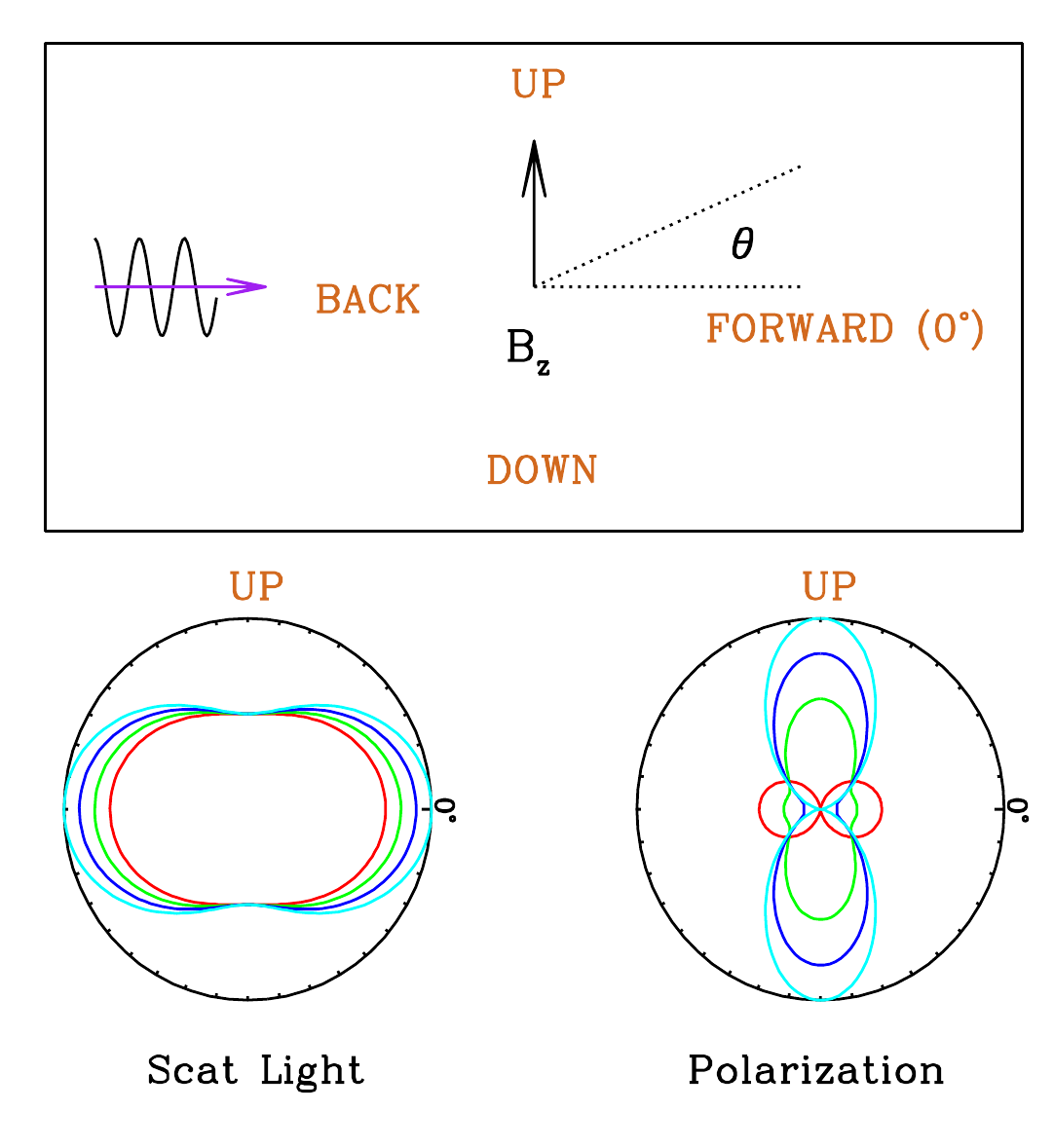}
\begin{minipage}{6in}
\centering
\caption{Top: Simple scattering geometry to illustrate the Hanle effect.
Unpolarized radiation is incident from the left on a scatterer where
there is a vertical magnetic field ($B_{\rm z}$).  Light is scattered
by angle $\theta$, with $0^\circ$ being forward, and other directions
as labeled.  Bottom: Scaled redistribution of scattered light and
polarization by the Hanle effect for 5 field strengths, from $B_{\rm
z}=0$ for magenta to the ``saturated'' limit for large $B_{\rm z}$
(red), shown in the form of polar plots.  Based on expressions
(26)--(28) from \cite{1999ApJ...520..335I}.
\label{f2}}
\end{minipage}
\end{figure}

\subsubsection{Zeeman Effect}

The Zeeman effect is well known for
producing line splitting of spectral lines in the presence of
magnetism.  However in astrophysical contexts, observed splittings
are rare, normally requiring
strong fields ($\sim$MG).  More typical are weaker fields leading
to linear and circular polarizations.  These arise the transverse and
longitudinal Zeeman effects, respectively
\citep[e.g.,][]{1994ASSL..189.....S}.  For modest
field strengths ($\sim$kG), the lines are not distinctly
split (although Zeeman broadening may be detected); instead,
polarization results from considerations of spectral line
gradients.  The transverse effect produces linear polarization and derives
from a second derivative of the Stokes-I profile shape and is 
proportional to the square of the field strength, $B^2$.  Easier
to measure is circular polarization from
the longitudinal effect, which is linear in $B$, derives
from a first derivative, and is antisymmetric about line center.

Detection of the Zeeman effect in massive stars is
challenging.  Line broadening tends to lessen the first derivative,
so that polarization scales as
$v_Z / v_{\rm broad}$.  Here $v_Z = c\,\Delta \lambda_Z/\lambda_0$
for the Zeeman effect in velocity units, with $\Delta \lambda_Z$ the Zeeman
shift for a line at $\lambda_0$.  Then $v_{\rm broad}$ may be
thermal, rotational, or wind broadening.  This challenge has been
successfully overcome for measuring surface stellar magnetism through
a co-adding procedure known as ``LSD'' to combine numerous spectral
lines and increase signal-to-noise as $\sqrt{N}$, for $N$ the number
of lines \citep[e.g.,][]{2009ARA&A..47..333D}.
One advantage of UV spectropolarimetry is it offers far more lines
as compared to the optical in order to measure magnetism in O stars,
in particular numerous iron lines \citep{2022Ap&SS.367..125F}.

\subsubsection{Hanle Effect}

This is another diagnostic of magnetism from spectral line formation.
It refers to an especially weak field limit in which the Zeeman
broadening has not fully lifted the degeneracy of the magnetic
sublevels, when the Zeeman splitting is
of order the natural line broadening.  In laboratory atomic physics,
one controls the field and infers level lifetimes (i.e., the inverse
of Einstein $A$-values).  In astrophysics, the situation is reversed.

The Hanle effect operates in resonance scattering
lines and leads to linear polarization effects.  In classical terms
the harmonic oscillator of the radiating dipole is caused to precess
at the Larmor frequency, $\omega_L$.  Consequently, magnetic
sensitivity is achieved on the order of $\omega_L \sim A$, which tends
to be 0.1--100~G fields, depending on the $A$-value, but typically
largerfield strengths 
for lines of shorter wavelength since $A \sim 1/\lambda^2$.

The Hanle effect has not been observed in massive stars, although
it has been successfully employed in solar studies
\citep[e.g.,][]{2013A&ARv..21...66S}.  The Hanle effect amounts to
redistribution of scattered light (see Fig.~\ref{f2}).  UV is
critical for massive star applications since that is 
where scattering resonance lines are found.  There are three
main points for use of the Hanle effect 
\citep{1997ApJ...486..550I}.  (1) The effect operates in lines
whether photospheric or circumstellar.  If circumstellar, the Hanle
effect allows to probe directly the magnetic field strength and
topology at levels that are difficult for the Zeeman effect.  (2)
Resonance lines have vastly higher cross-sections than electron
scattering.  For targets with low optical depth circumstellar media
in electron scattering, resonance line polarization may still be
significant.  (3) A multiline approach is key for extracting
information about the magnetism.  For example, not all lines are Hanle
sensitive.  Common UV resonance lines for massive stars are Li-like
doublets for which one component produces no resonance polarization,
nor Hanle effect.  That component serves as a control to infer magnetic
detection using the Hanle
sensitive one.

\subsubsection{Dilution Effect}

\begin{figure}[t]
\centering
\includegraphics[width=4.25in]{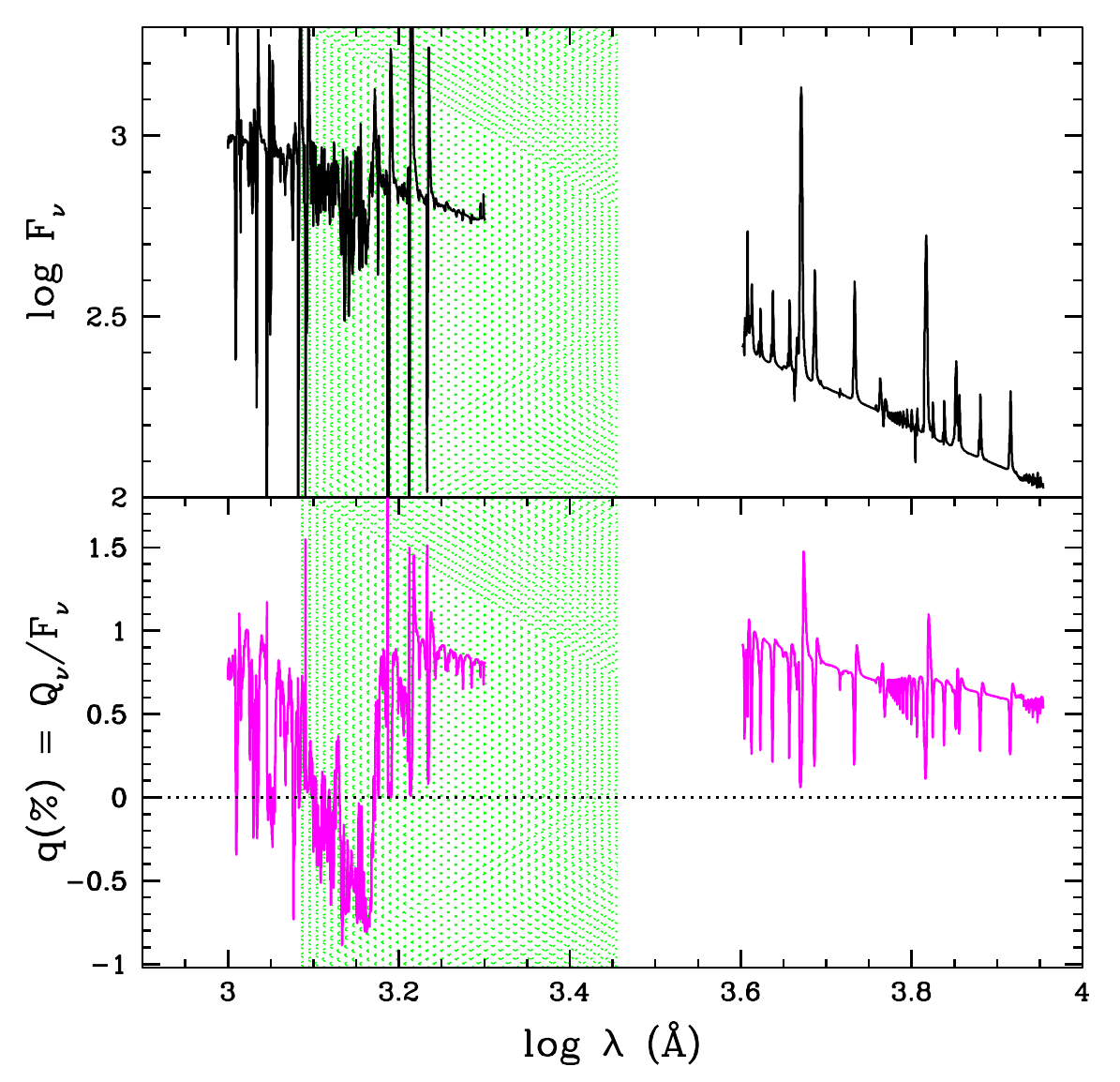}
\begin{minipage}{6in}
\centering
\caption{Synthetic spectropolarimetry for a WR~star using
CMFGEN \citep{1994A&A...289..492H}.  The wind is axisymmetric with
an equatorial density of $3.3\times$ great than the pole.  Top
is a relative flux; bottom is the linear polarization $q$, where
$Q_\nu$ is the polarized flux.  Note the
dilution effect in numerous lines for both UV and optical.
The change in sign for the UV arises from iron-line blanketing
and signifies a $90^\circ$ position angle (PA) change.  
Green shading is the {\em Polstar} waveband.
\label{f3}}
\end{minipage}
\end{figure}

Winds and disks of massive stars show spectral lines that tend
to be resonance or recombination.  These lines can form over vastly
different volumes of the circumstellar environment, subject to
ionization distribution and detailed NLTE level populations.
In some cases a particular spectral transition may form in a region
where line photons are not much scattered by the electrons. 
Traditionally, polarization
is measured as a ratio of polarized flux to total flux.
When the line photons are not scattered, they contribute to the
total flux and reduce the polarization owing to higher normalization
relative to the neighboring continuum (see examples
in Fig.~\ref{f3}).  This is the
``dilution effect'', because the polarization flux will not have
changed yet the total flux has
\citep[e.g.,][]{1998MNRAS.296.1072H}.  

The dilution effect is useful
for two reasons:  first it demonstrates the presence of intrinsic
source polarization, because the interstellar polarization is not
subject to dilution. Second, the dilution
may be so severe as to yield a measure of the interstellar
polarization at that wavelength.  There is in general no reason for
the continuum and interstellar polarizations to have the same 
position angle (PA),
so severe dilution of polarization may be attended by an observed
rotation in PA across the emission line due to the interstellar
contribution.

\subsubsection{Line Absorption Effect}

Complex circumstellar structures can lead to differential absorption
effects by line opacity in relation to where scattering polarization
in the continuum is formed.  There are two main opportunities of
interest.  First is line blanketing.  In a study of the interacting
binary $\beta$~Lyr combining UV continuum polarization from WUPPE
with ground-based optical polarization from HPOL, the polarization
is seen to rotate by 90 degrees \citep{1998AJ....115.1576H}.  A PA
rotation is a clear signature for a change in geometry.  
For $\beta$ Lyr the PA rotation began shortward of the
Balmer jump and persisted into the UV.  The former is indicative
of higher absorbing bound-free opacity in the accretion disk of the
system; the continuation into the UV from a forest of 
absorbing iron lines .
However, the polarization was not eliminated, only rotated, and suggested
a bipolar jet outflow from the disk, consistent with  
with evidence from optical interferometry
\citep{1996A&A...312..879H}.  A different
example of blanketing is seen in Figure~\ref{f3}, now for an
axisymmetric WR wind with $3.3\times$ higher density at the equator
than the pole.  The sign change in polarization at UV wavelengths
arises from iron line blanketing, with net polarization coming from
scattering by the less opaque polar wind.

Our second example pertains to differential absorption from complex
structures in a cyclic fashion.  The WR star WR~6 shows variation in
polarization across the strong He{\sc ii} 4686 emission line.  The
variation is peculiar in that polarization is reduced over redshifted
velocities showing a linear feature in a $q-u$ diagram, whereas
a loop is traced for blueshifted velocities (a ``blue loop'').  The
polarization for the redshifted portion of the line follows the
standard dilution effect.  The blue loops signify something else.

WR~6 is a candidate for a co-rotating interaction region
\citep[CIR;][]{2018MNRAS.474.1886S}.  A CIR is a spiral structure
threading a wind owing to the presence of a brightness enhancement at
the stellar surface.  The enhancement drives a faster wind leading
to a shock and post-shock cooling region, which in the co-rotating
frame takes on the shape of a spiral.  For an outflowing wind,
\cite{2023MNRAS.526.1298I} demonstrated that such a structure could
indeed account for the blue loop, for those rotational phases when
the CIR is passing in front of the stellar photosphere with limb
polarization.  {\em Polstar} will provide access to a rich spectrum
of UV lines combining velocity shift information with polarimetry
as demonstrated for WR~6.

\begin{figure}[t]
\centering
\includegraphics[width=3in]{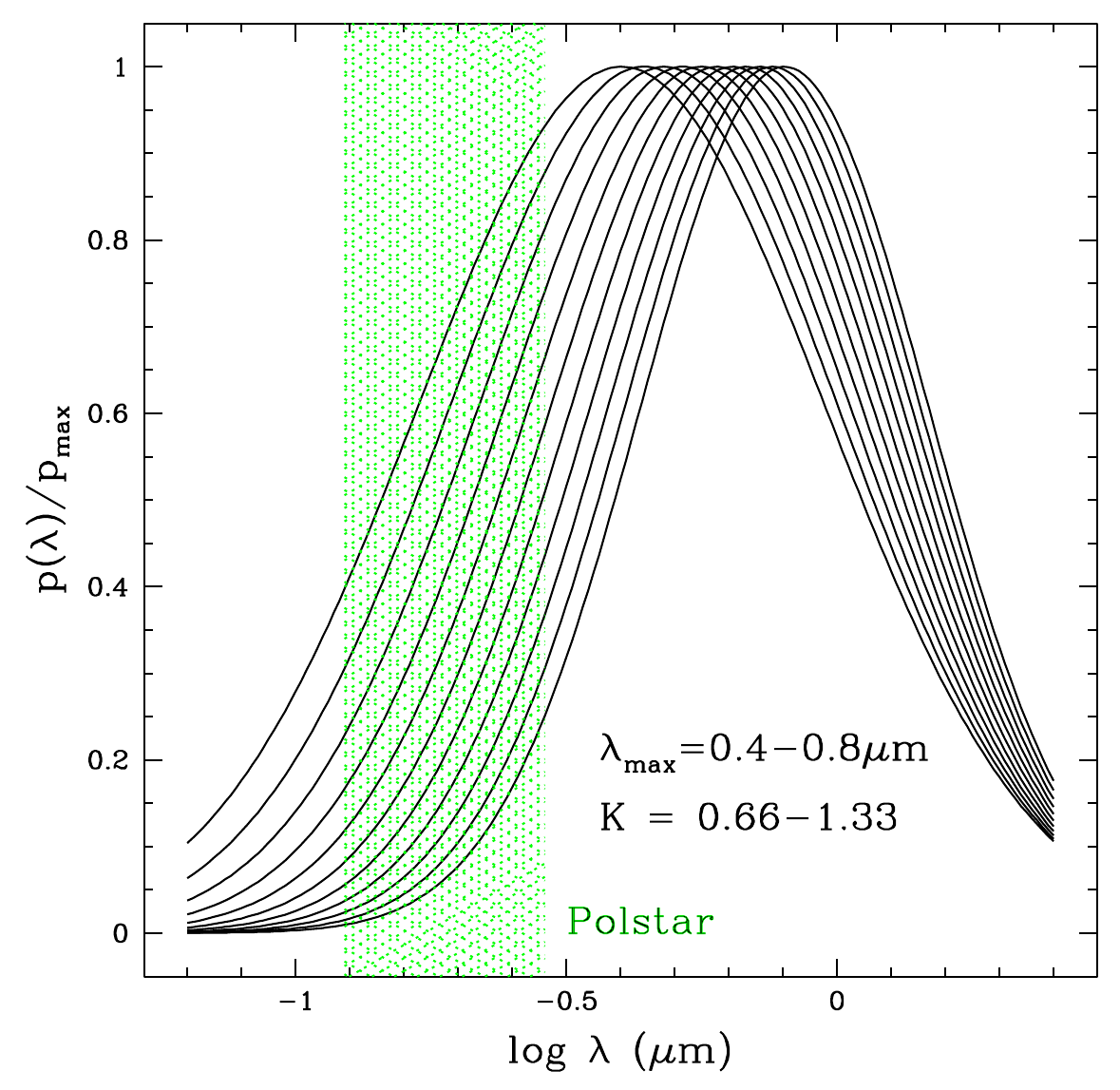}
\includegraphics[width=3in]{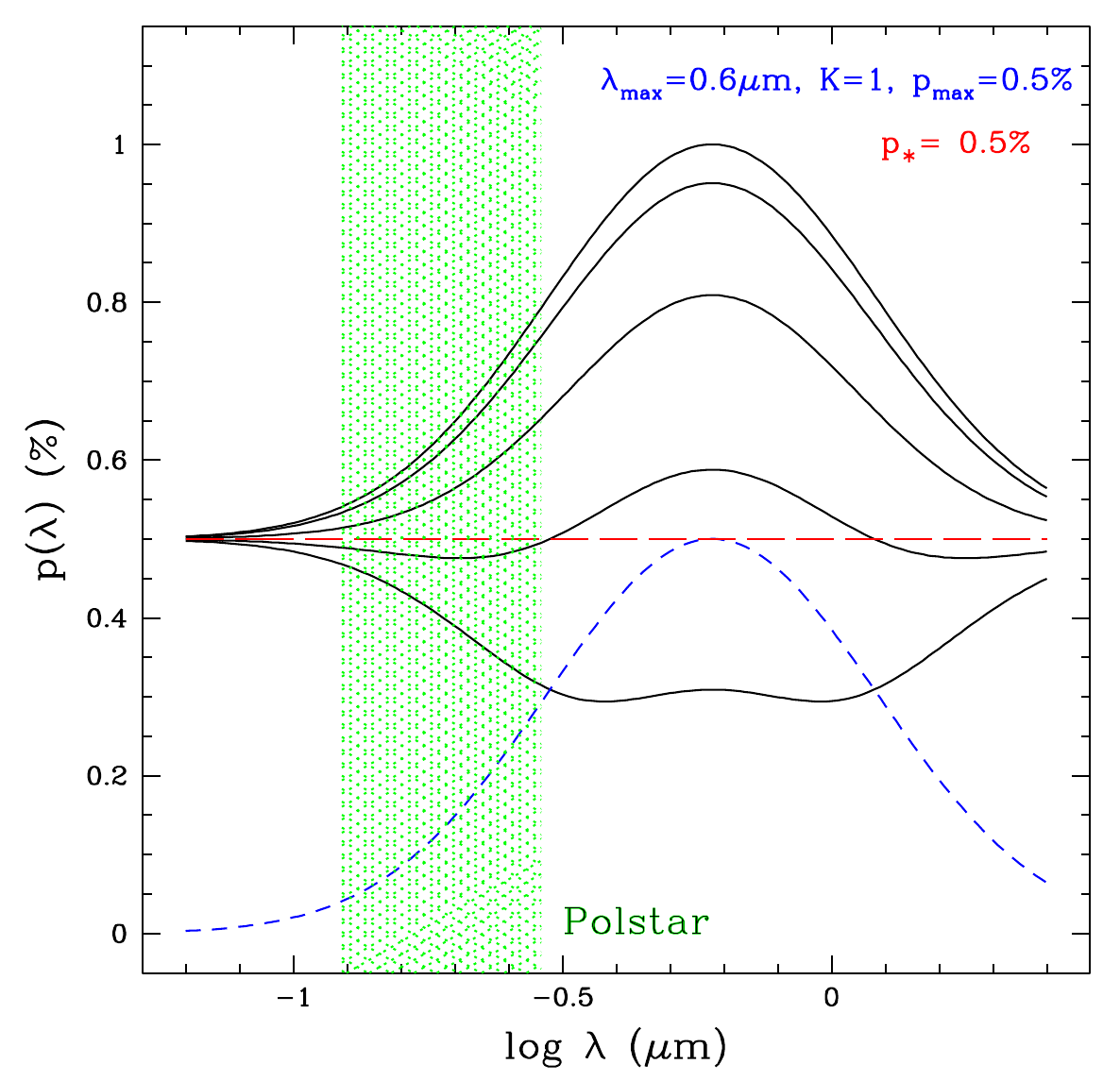}
\begin{minipage}{6in}
\centering
\caption{Left:  Several Serkowski Law curves plotted for a range
of $\lambda_{\rm max}$ and $K$ values as indicated.  The polarization
is normalized to $p_{\rm max}$.  The green hatch is the {\em Polstar} 
waveband, where interstellar polarization is rapidly dropping.
Right:  Illustration for how Thomson scattering and interstellar
polarization could combine.  Blue is the Serkowski Law; red is
the stellar polarization.  Black curves are combinations in which
the PA between the two differ from $0^\circ$ to $90^\circ$
in $18^\circ$ increments.
\label{f4}}
\end{minipage}
\end{figure}

\subsection{Interstellar Polarization}

As starlight passes through the interstellar medium, it
obtains a linear polarization even if the star is spherical
This arises from interstellar magnetism aligning dust
grains, which then absorb a preferential sense of
polarized light, thereby acting as an absorptive filter leaving a
residual polarization orthogonal to the direction of grain alignment.  The
interstellar medium thus imposes a polarization of some
amplitude at fixed PA, but which is wavelength-dependent.
The form of this polarization is well-modeled by the Serkowski Law
\citep{1975ApJ...196..261S}, with the following form:

\begin{equation}
p_\lambda = p_{\rm max}\,e^{-K\,\ln^2(\lambda_{\rm max}/{\lambda})},
\end{equation}

\noindent where $p_{\rm max}$ is the maximum polarization at
$\lambda=\lambda_{\rm max}$, and $K$ controls the width of the
curve.  For a reasonably large sample, $K \approx 1.66\, \lambda_{\rm
max}(\mu m)$ with $\lambda_{\rm max}$ ranging from 0.4 to 0.8 $\mu
m$ \citep[e.g.,][]{1992ApJ...386..562W}.

Example plots of the Serkowski Law are shown in Figure~\ref{f4},
left side.  These are for a range of $\lambda_{\rm max}$ values,
with $K$ determined by the relation above.  The curves are normalized
to $p_{\rm max}$.  The right panel figure combines a flat stellar
polarization from Thomson scattering.  The different curves result
from changing the PA between the star and interstellar, as explained
in the caption.  Observing in the UV with {\em Polstar} has the
advantage that interstellar polarization is dropping and becoming
far less confounding to stellar studies.

\section{The Polstar Science Theme}

\begin{table}
\begin{minipage}[t]{.40\textwidth}
  \centering
\caption{{\em Polstar} Attributes \\ 
	\label{t2}}
\begin{tabular}{ll}
\hline Aperture	& 40~cm \\
$A_{\rm eff}$ @150 nm & 22 cm$^2$ \\
Waveband (Spec)	& 115--286 nm \\
Waveband (Pol)	& 122--286 nm \\
Resolving Power	& 20,000$^a$ \\ \hline
\end{tabular}
\centerline{\small $^a$ Corresponds to 15 km/s velocity resolution.}
\end{minipage}\qquad
\begin{minipage}[t]{.60\textwidth}
  \centering
\caption{Instrument Characteristics\\ \label{t3}}
\begin{tabular}{lcr}
\hline Subexposure & $\delta t$ & 2~s-100~s \\
Exposure & $t_{\rm exp}$ & $6\times\delta t$ \\ 
Source Count Rate & $\dot{S}_\ast$ & counts/s \\
Source Counts & $S_\ast$ & $\dot{S}_\ast\,\delta t$ \\
Source Noise & $N_\ast$ &  $\sqrt{S_\ast}$ \\
Read Noise & $N_R$ & 5.4/subexposure \\
Dark Current & $\dot{S}_D$ & 0.06 Hz/pix \\
Dark Noise & $N_D$ & $\sqrt{\dot{S}_D \delta t}$ \\ \hline
\end{tabular}
\end{minipage}
\end{table}

The {\em Polstar} mission concept involves a UV spectropolarimeter
as summarized in Table~\ref{t2}.  The telescope will have a
40~cm aperture with an effective area of about 22~cm$^2$ at a fiducial
wavelength of 150~nm (near the important C{\sc iv} doublet).
Spectroscopy will be available from 115 to 286~nm, providing for
observations from Ly$\alpha$ to the Mg{\sc ii} doublet.  Polarimetric
sensitivity ends at 122~nm.  
Spectropolarimetry will be achieved from
the N{\sc v} doublet to Mg{\sc ii}, to include numerous
important diagnostic lines that will be observed at $R=20000$.

\subsection{Science Objectives}

The goal of the {\em Polstar} mission is to understand better the
lives of massive stars and their impacts on galaxy evolution.  The
theme is closely aligned with many priorities of the current and
recent Decadal Surveys and the NASA Astrophysics Roadmap.  Specifically
we seek to understand angular momentum evolution in massive stars.
Essentially all massive stars are born into binaries (or triples,
and even higher orders).  Moreover, nearly all massive multiples
will exchange mass at some time,
often short-lived yet crucially important phases.  Indeed, mergers
are expected in some cases.
The interactions imply not only mass transfer (whether
conservative or non-conservative) but attendant exchanges in
angular momentum.  Issues of angular momentum can lead to spin-up
or spin-down of stars, and govern the evolution of binary orbits.

The {\em Polstar} team has developed three main science categories 
to be addressed with UV spectropolarimetry:

\begin{enumerate}

\item We seek to understand better the origin of rotational
characteristics observed in massive stars.  Several of the brightest
massive stars in the sky are in fact in near-critical rotation
\citep[e.g., Achernar, $\zeta$~Pup, and $\epsilon$~Sgr, to name a few;][]
{2003A&A...407L..47D, 2024MNRAS.529..374B, 2024arXiv240711352B}.

\item A better understanding of how rapid rotation affects the surfaces
of massive stars.  Here rapid need not be critical.  Critical rotation
leads to distortion of the isopotential of the star,
accompanied by polar brightening and equatorial darkening.
However, many massive stars are rotating only around 30\% of critical,
sufficient to induce mixing which is predicted (and observed) to
enhance surface nitrogen abundances \citep[e.g.,][]{2008ApJ...676L..29H}.
Those that are near-critical rotating are often correlated with having
circumstellar disks, such as the Be~stars, yet not in all cases such as the
the fast rotating Bn~stars.

\item Finally, {\em Polstar} will address processes that regulate
the transfer of angular momentum.
For example, around 10\% of massive
stars have surface magnetism at the kG~level.  Many of these have
suffered significant magnetic braking.  Separately, the Be~stars may
all have remnant companions from mass transfer (subdwarf O and B
stars, white dwarfs, and even neutron stars or black holes)
\citep{2019ApJ...885..147K}, and
the Be~stars are certainly fast (perhaps near-critical) rotators.
The Be~stars also have decretion
disks, but these seems to come and go, with the star able to eject
gas from its surface, and place it into orbits with higher specific
angular momentum.  {\em Polstar} will clarify how individual stars
transport angular momentum throughout their interiors and in
interactions with their immediate environments.

\end{enumerate}

\begin{figure}[t]
\centering
\includegraphics[width=3in]{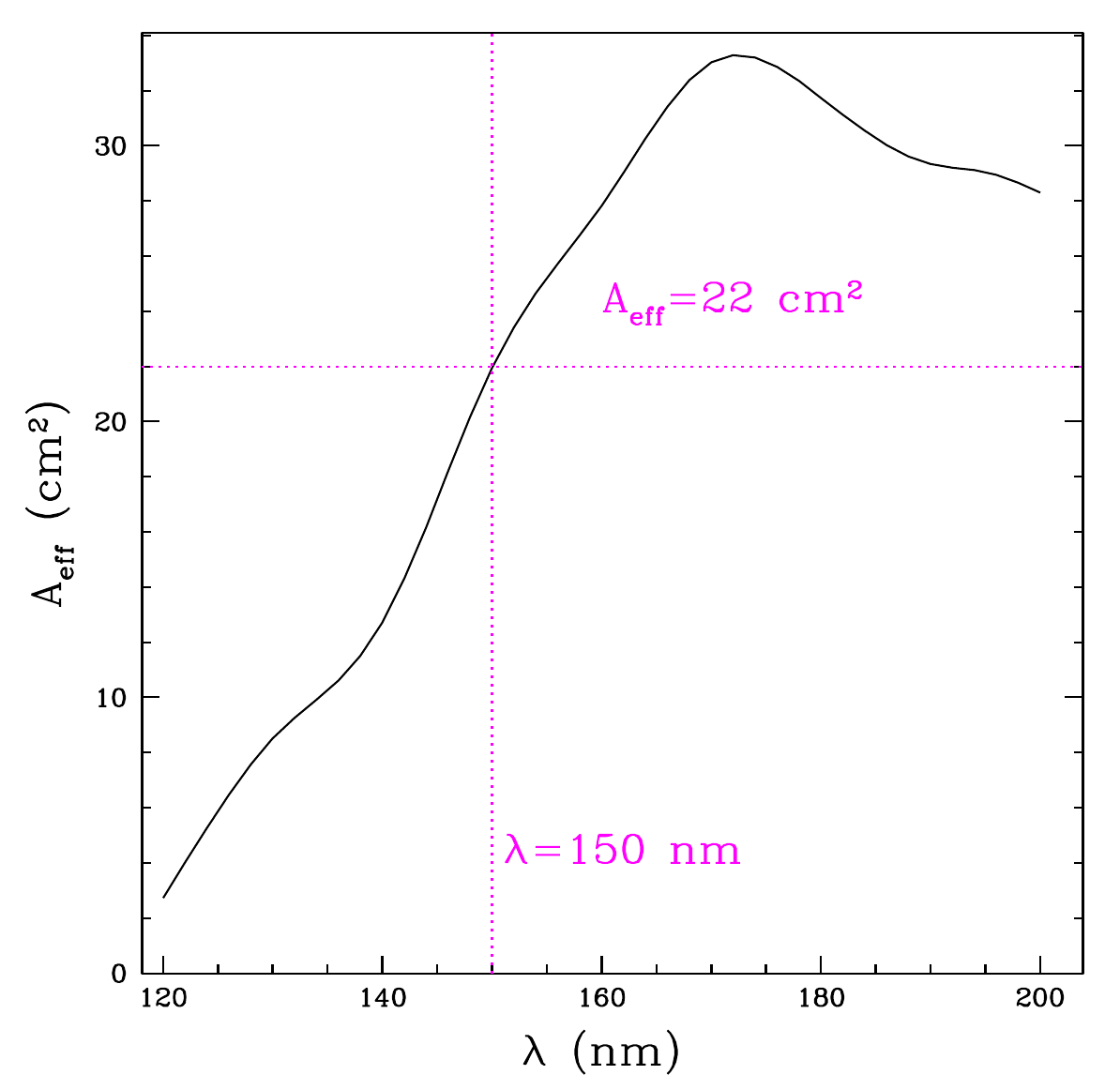}
\includegraphics[width=3in]{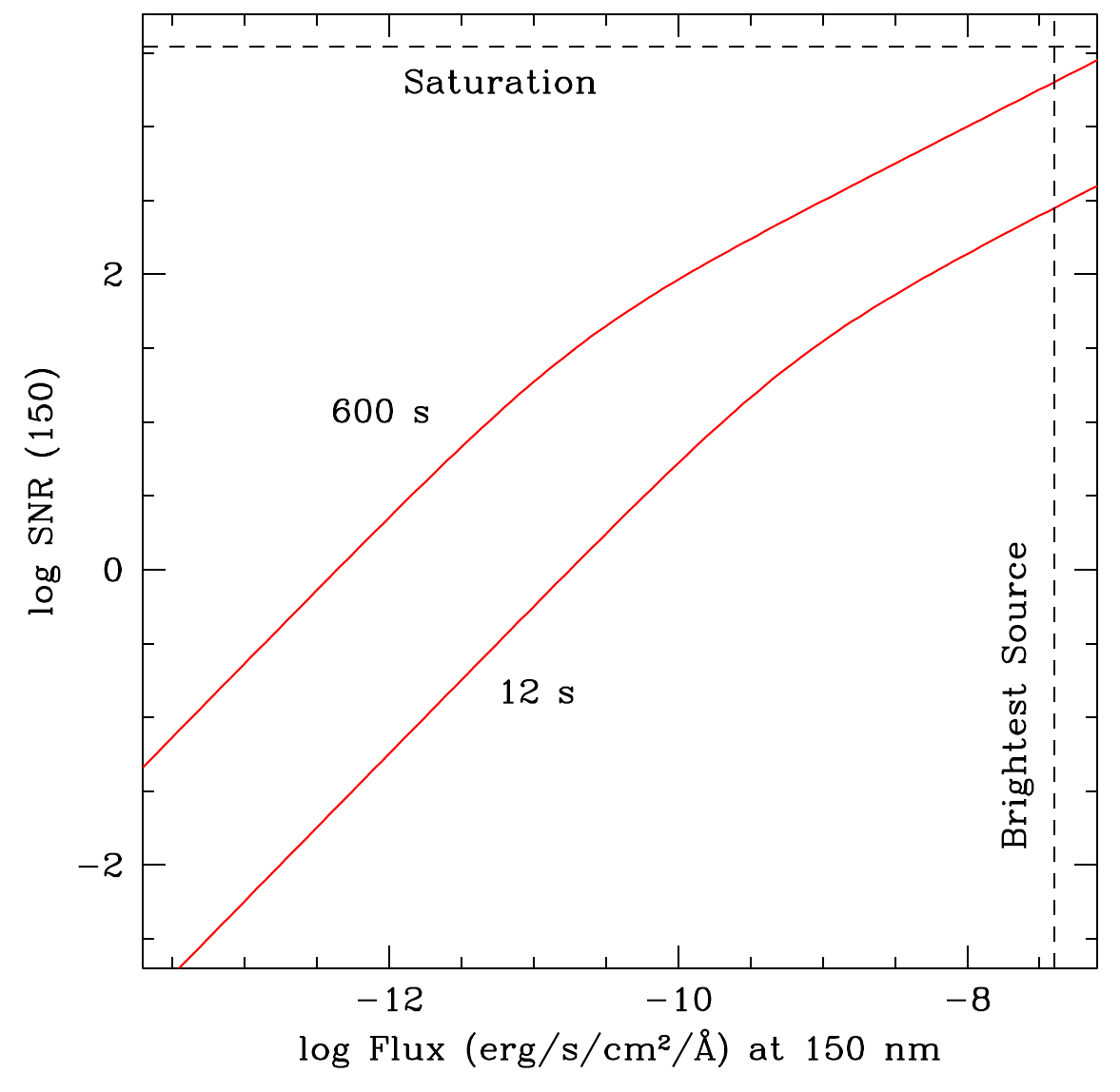}
\begin{minipage}{6in}
\centering
\caption{Left:  Anticipated effective area of {\em Polstar}.
Right:  SNR values at 150~nm vs source flux.  The two red curves
are for an exposure of 12 and 600~s.  The vertical dashed line
is for our brightest source, Rigel.  The horizontal dashed line
is for saturation.
\label{f5}}
\end{minipage}
\end{figure}

\subsection{Polstar Observations}

The {\em Polstar} design will make use of a Wollaston prism to split
the light beam into ordinary and extraordinary rays to an echelle.
Measurements will be obtained at 6 different modulator positions
at selected orientation angles in order to optimize instrumental
performance.  The brightness throughput of the assembly depends on
the polarization of the beam, and the 6 measurements can be used
to invert (with optimization) to achieve full Stokes spectropolarimetry
in I, Q, U, and V \cite[refer to][]{2022Ap&SS.367..121S}.  We adopt
the language that data taken at each modulator position is a
subexposure; the 6 subexposures together are a single exposure;
each visit to a target is an observation which may consist of
multiple exposures plus all associated overtimes.  As a concrete
example, consider spectropolarimetry of a binary system with known
period.  Imagine data are collected at 20 equally spaced orbital
phases, each visit having 2 exposures.  This program would then
consist of 20 observations, with 40 exposures, and 240 subexposures.
There would be 240 total-light spectra and 40 polarized spectra
produced.  The subexposure time (for spectroscopy) or exposure time
(for polarimetry) would be dictated by a signal-to-noise calculation
in relation to the science requirements.

Using a model of Poisson statistics, we estimate the signal-to-noise
ratio (SNR) for each spectral resolution element (hereafter
``resel'').  Each of the 6 orientation angles produces a spectrum,
which are combined to obtain the polarization measures Q,
U, and V.  We estimate the SNR in terms of subexposures, from which
SNR values for exposures, observations, and programs can be found.
Refer to Table~\ref{t3} for definitions of symbols appearing in the
expressions that follow.

The signal is the source count rate multiplied by the subexposure time.
The total noise, $N_T$, consists of source noise, dark
current, and read noise, combined in quadrature as

\begin{equation}
N_T^2 = N_\ast^2 + N_D^2 + N_R^2 = \dot{S}_\ast\delta t 
	+ \dot{S}_D \delta t + N_R^2
\end{equation}


\noindent The resultant SNR in a subexposure becomes

\begin{equation}
SNR_{\rm sub} = \sqrt{\frac{\dot{S}_\ast^2\,\delta t^2}{\dot{S}_\ast\, \delta 
        t + \dot{S}_D \,\delta t + N_R^2}} .
\end{equation}

\begin{figure}[t]
\centering
\includegraphics[width=5.5in]{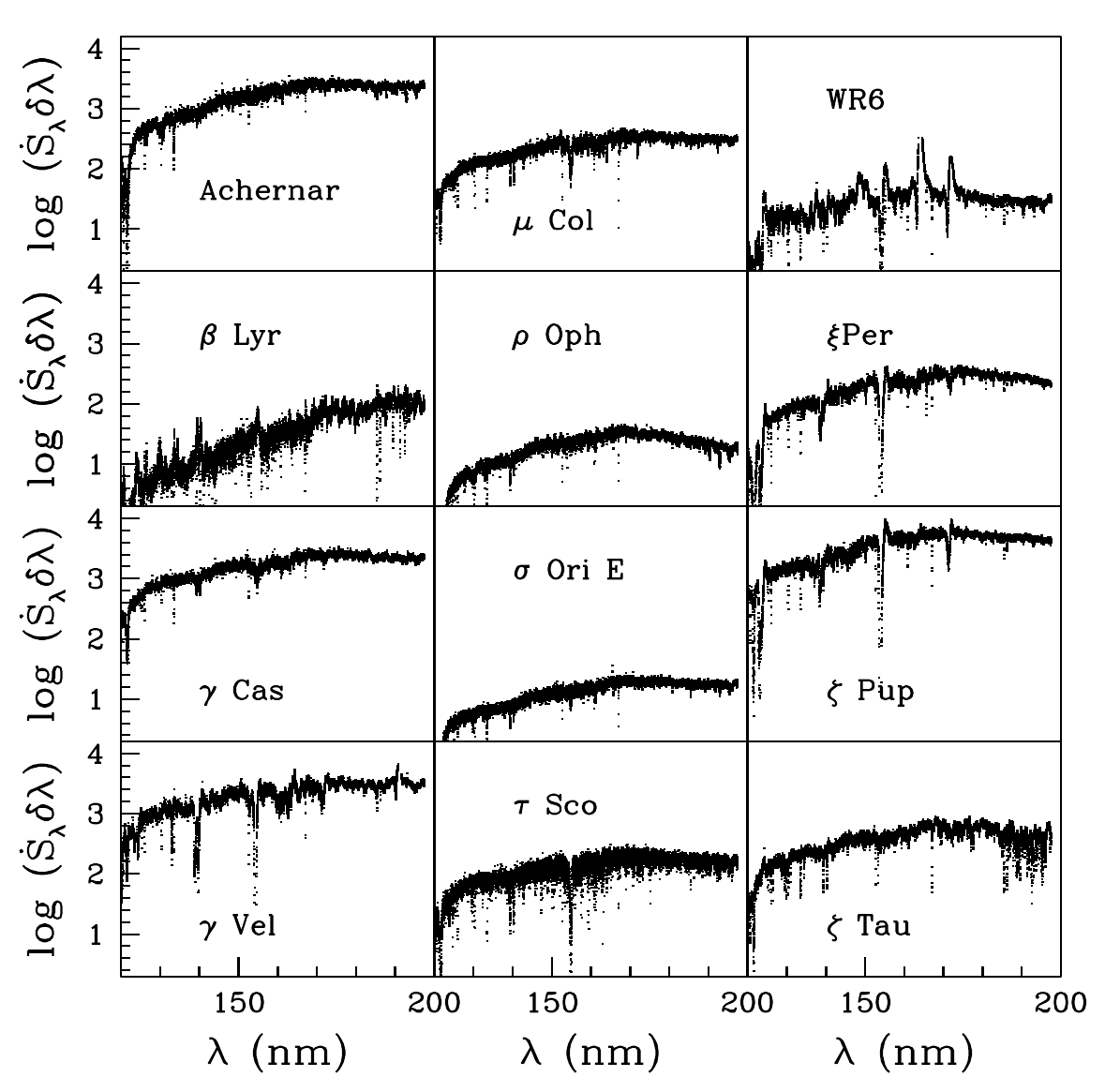}
\begin{minipage}{6in}
\centering
\caption{Simulated count rate spectra , 
involving the resel wavelength bin $\delta \lambda = \lambda/R$,
for a selection of well-known massive stars
using archival IUE spectra.
\label{f6}}
\end{minipage}
\end{figure}

\noindent As an example, the source count rate can be determined for
known flux, using

\begin{equation}
\dot{S}_\ast = \frac{\lambda\,f_\lambda}{hc}\times A_{\rm eff}(\lambda)\times
        \frac{\lambda}{R},
\end{equation}

\noindent where $f_\lambda$ is the source specific flux and $A_{\rm
eff}(\lambda)$ is the effective area.  Figure~\ref{f5} shows the
effective area with wavelength (left) and SNR performance as a
function of $f_\lambda$ at a reference wavelength of 150~nm (right).
Additionally, Figure~\ref{f6} provides example count rates per resel
for some well-known massive stars using SWS IUE spectra.

{\em Polstar} will provide spectropolarimetric data of massive
stars.  Binning and stacking of exposures will provide high, medium,
and low resolution spectroscopy with time to achieve diverse
scientific requirements.  For polarization many interesting effects
arise at the level of 0.1\%, requiring $SNR$ values in excess of
1000 at the spectral resolution of interest.  {\em Polstar} is
baselined to provide polarimetric precision down to 0.03\% in the
most favorable cases.

\section{Summary}

Polarization is a powerful diagnostic for geometry in unresolved
sources.  The {\em Polstar} UV spectropolarimetry mission would
provide sorely needed new capability to address several
outstanding and fundamental questions that undergird massive
star evolution, namely the exchange of angular momentum between
stars and its internal transport within them.  A Guest 
Observer program is anticipated to
provide important datasets in other areas of astrophysics
(e.g., novae, chromospheres, exoplanets, and more).



\begin{acknowledgments}
RI gratefully acknowledges support from the National Science
Foundation under grant number AST-2009412.  Figures~\ref{f1} and
\ref{f3} are based on calculations kindly provided by J.~P.~Harrington
and D.~J.~Hillier, respectively.  The authors are indebted to the
entire {\em Polstar} Team for their efforts past and present with
developing and promoting the UV spectropolarimetry mission concept.
\end{acknowledgments}

\begin{furtherinformation}

\begin{orcids}

\orcid{0000-0002-7204-5502}{Richard}{Ignace}
\orcid{0000-0002-9071-6744}{Paul}{Scowen}

\end{orcids}


\begin{conflictsofinterest}
The authors declare that there are no conflicts of interest.
\end{conflictsofinterest}

\end{furtherinformation}



%

\bibliographystyle{bullsrsl-en}


{\small 
\bibliography{r_ignace}
}

\end{document}